\documentclass[fleqn,twoside]{article}
\usepackage{espcrc2}
\usepackage{axodraw}
\usepackage{epsfig}
\readRCS
$Id: espcrc2.tex,v 1.2 2004/02/24 11:22:11 spepping Exp $
\usepackage{graphicx}
\usepackage[figuresright]{rotating}

\newcommand{\AmS}{{\protect\the\textfont2
  A\kern-.1667em\lower.5ex\hbox{M}\kern-.125emS}}

\title{Some recent results on evaluating Feynman integrals}

\author{V.A.~Smirnov\address[MCSD]{
Nuclear Physics Institute of Moscow State University,\\
Moscow 119992, Russia}%
        \thanks{Supported by
the Russian Foundation for Basic Research through grant
05-02-17645.
To appear in the proceedings of the
7th International Symposium on Radiative Corrections (RADCOR05),
Shonan Village, Japan, 2005.}}

\newcommand{\be}{\begin{equation}}
\newcommand{\ee}{\end{equation}}
\newcommand{\bea}{\begin{eqnarray}}
\newcommand{\eea}{\end{eqnarray}}

\newcommand{\Gm}{\Gamma}

\newcommand{\ep}{\epsilon}

\newcommand{\lm}{\lambda}

\newcommand{\sg}{\sigma}

\newcommand{\pa}{\partial}
\newcommand{\dd}{\mbox{d}}

\newcommand{\nn}{\nonumber}

\begin{document}

\begin{abstract}
Some recent results on evaluating Feynman integrals are
reviewed. The status
of the method based on Mellin-Barnes representation as a powerful
tool to evaluate individual Feynman integrals is characterized.
A new method based on Gr\"obner bases to solve integration by parts
relations in an automatic way is described.
\end{abstract}

\maketitle

\section{Introduction}

Perturbative quantum-theoretical amplitudes are
expressed in terms of Feynman
integrals over loop momenta. Usually, one turns immediately to
scalar Feynman integrals using some projectors
and obtains a family of scalar Feynman integrals
with the same structure of the integrand and
various powers of propagators (indices):
\bea
F(a_1,\ldots,a_n) &=&
\int \cdots \int \frac{\dd^d k_1\ldots \dd^d k_h}
{E_1^{a_1}\ldots E_n^{a_n}}\,,
\label{eqbn}
\eea
where $k_i$, $i=1,\ldots,h$, are loop momenta
and the denominators $E_r$ are either quadratic of linear with
respect to  $k_i$ and external momenta
$q_1,\ldots,q_N$.
By default, the integrals are dimensionally regularized
with $d=4-2\ep$.

If the number of Feynman integrals needed for a given calculation
is small or/and they are simple, one evaluates, by some methods,
every scalar Feynman integral of the given family.
Various methods are used,
in particular, alpha and/or Feynman parameters,
Mellin--Barnes (MB) representation \cite{MB1,MB2}
and differential equations \cite{DE}.
In the next section, the method of  MB representation is briefly
reviewed.

If it is necessary to evaluate a lot of complicated
Feynman integrals (\ref{eqbn}) the standard way
is to apply integration by parts (IBP) \cite{IBP} relations
\bea
\int\dd^d k_1\int \dd^d k_2 \ldots
\frac{\pa}{\pa k_i}\left( p_j
\frac{1}{E_1^{a_1}\ldots E_n^{a_n}}
\right)   =0 \nn
\eea
in order to construct an algorithm that gives and expression
of a given Feynman integral as a linear combination of
some {\em master} integrals,
$F(a_1,\ldots,a_n)=\sum_i c(a_1,\ldots,a_n) I_i$
Here  $p_1=k_1,\ldots,
p_h=k_h, p_{h+1}=q_1,\dots, p_{h+N}=q_N$.
Once one has a solution of this (reduction)
problem, it becomes sufficient to evaluate only master
integrals.
The IBP relations can be written as $f_i\cdot F$
where $f_i$ are polynomials in the shift operators
$Y^{\pm 1}_i$ defined by 
$(Y^{\pm 1}_i\cdot F)(a_1,\ldots,a_n)=F(a_1,\ldots,
a_i+1,\ldots,a_n)$.

The first attempt to make the reduction procedure
systematic was based on the fact that the total number
of IBP
equations grows faster than the number of Feynman
integrals satisfying the condition $\sum_i |a_i|\leq M$ at
$M\to \infty$ so that, at sufficiently large $M$, one obtains an
overdetermined system of equations \cite{LGR1,LGR2} which
can be solved. There is already a public implementation of this
algorithm on a computer \cite{AnLa}.
Another attempt \cite{Bai} is based on a
special parametric representation.

Other attempts are based on the use of
Gr\"obner bases (GB) \cite{Buchberger}. This idea was first suggested in
\cite{Tar1}, where IBP relations were reduced to differential
equations. To do this, it is assumed that there is a non-zero mass
for each line. The typical combination $a_i Y_i$, is then
naturally transformed into the operator of differentiation in the
corresponding mass.
An attempt to use GB associated with the shift operators
was made in \cite{Gerdt}.
However, the corresponding algorithms
can now work only in simplest cases,
for $n=2$.

In Section~3, another approach \cite{2S} based on GB
is briefly characterized.
In conclusion, I shall discuss some perspectives.

\section{Evaluating by MB representation}

The method of MB representation was successfully applied
to evaluate massless on-shell
double \cite{MB1,MB2,GTGR0ATT,SV,AGORT} and triple \cite{3b,BDS}
boxes, with results written in terms of harmonic polylogarithms
\cite{HPL}, double boxes with one leg off shell \cite{LGR2,S2}
and massive on-shell double boxes \cite{HS,CGR}
The method is based on the MB representation
\bea
\frac{1}{(X+Y)^{\lm}} = 
\int_{- i  \infty}^{+ i  \infty}
\frac{Y^z}{X^{\lm+z}} \frac{\Gm(\lm+z) \Gm(-z)}{\Gm(\lm)}
\frac{\dd z}{2\pi  i}
\nn
\eea
applied to replace a sum of two terms raised
to some power by their products in some powers.

Experience shows that a minimal number of MB integrations
is achieved if one introduces MB integrations loop by loop,
i.e. derives a MB representation for a one-loop subintegral,
inserts it into a higher two-loop integral, etc.
For example, for the tennis court graph
shown in Fig.~1,
\begin {figure} [htbp]
\begin{picture}(130,70)(-40,18)
\Line(-15,0)(0,0)
\Line(-15,100)(0,100)
\Line(115,0)(100,0)
\Line(115,100)(100,100)
\Line(0,0)(50,0)
\Line(50,0)(100,0)
\Line(100,50)(50,50)
\Line(50,50)(0,50)
\Line(0,100)(0,0)
\Line(50,0)(50,50)
\Line(100,0)(100,100)
\Line(0,100)(100,100)
\Vertex(0,0){1.5}
\Vertex(50,0){1.5}
\Vertex(100,0){1.5}
\Vertex(0,50){1.5}
\Vertex(50,50){1.5}
\Vertex(100,50){1.5}
\Vertex(100,100){1.5}
\Vertex(0,100){1.5}
\Text(-22,0)[]{$p_2$}
\Text(124,0)[]{$p_4$}
\Text(-22,100)[]{$p_1$}
\Text(124,100)[]{$p_3$}
\Text(25,43)[]{\small 1}
\Text(-7,25)[]{\small 2}
\Text(25,7)[]{\small 3}
\Text(75,7)[]{\small 4}
\Text(43,25)[]{\small 7}
\Text(93,25)[]{\small 5}
\Text(75,43)[]{\small 6}
\Text(50,93)[]{\small 8}
\Text(90,75)[]{\small 10}
\Text(-7,75)[]{\small 9}
\end{picture}
\end{figure}
one can start from a MB representation for a lower box subintegral
with three legs off shell, then insert it
into the double-box subintegral to obtain a
MB representation for the double box with two (upper) legs off
shell, insert it into the whole graph and obtain an eightfold MB
representation for the given family of integrals
with general powers of the propagators and
the power of the numerator
$[(l_1+ l_3)^2]^{-a_{11}}$,
where $l_{1,3}$  are the momenta flowing through lines
$1$ and $3$ in the same direction:
\bea
T(a_1,\ldots,a_{11};s,t;\ep)
&& \nn \\ &&  \hspace*{-36mm}
= \frac{\left(i\pi^{d/2} \right)^3 (-1)^a(-s)^{8-a-3\ep}}{
\prod_{j=2,4,5,6,7,8}\Gm(a_j) \Gm(4-a_{4567}-2\ep)t^2}
\nn \\ &&  \hspace*{-36mm}\times
\frac{1}{(2\pi i)^8} \int_{-i\infty}^{+i\infty}
\dd w \, \prod_{j=1}^7 \dd z_j
\left(\frac{t}{s} \right)^{w} \! \prod_{j=2}^7 \Gm(-z_j)
\nn \\ &&  \hspace*{-36mm} \times
\frac{\Gm(a - 8 + 3 \ep +w)
\Gm(8 - a - 3 \ep - w) }
{ \Gm(a_1 - z_2) \Gm(a_3 - z_3) \Gm(a_9 - z_6)\Gm(a_{10} - z_{47})}
\nn \\ &&  \hspace*{-36mm}
\times\frac{
\Gm(a_5 + z_{14}) \Gm(a_2 + z_{56})\Gm(2 - w + z_5)}
{\Gm(4 - a_{123} - 2 \ep + z_{123}) \Gm(8 - a -4 \ep - z_5)}
\nn \\ &&  \hspace*{-36mm}
\times
\frac{ \Gm(2 - a_{457} - \ep - z_{13})\Gm(a_{10} - 2 + w - z_{457}) }
{\Gm(a_{1234567,11} - 4 + 2 \ep + z_{4567})}
\nn \\ &&  \hspace*{-36mm}
\times
\Gm(2 - a_{567} - \ep - z_{124})
\Gm(a_{4567} - 2 + \ep + z_{1234})
\nn \\ &&  \hspace*{-36mm}
\times
\Gm(2 - a_{23} - \ep + z_{13} - z_5)  \Gm(a_9-2 + w - z_{56})
\nn \\ &&  \hspace*{-36mm}
\times
\Gm(2 - a_{12} - \ep + z_{12} - z_{567}) \Gm(a_7 + z_{123})
\nn \\ &&  \hspace*{-36mm}
\times
\Gm(a_{123} - 2 + \ep - z_{123} + z_{567})
\Gm(z_{57} - z_1)
\nn \\ &&  \hspace*{-36mm}
\times
\Gm(4 - a_{89,10} - \ep - w + z_{4567})  \, ,
\nn
\eea
with $a_{12}=a_1+a_2, a=a_{1\ldots,11}, z_{14}=z_1+z_4$,
etc.

Deriving MB representations for general indices is very useful
because one can apply it for various partial cases and
obtain crucial checks. For example, this very
representation was used in \cite{BDS} to calculate
$T(1,\ldots,1,-1)$, in a Laurent expansion in $\ep$ and to check
cross order relations in $N=4$ SUSY gauge
theories \cite{SUSY}.
Let us stress that one can check such a cumbersome representation
in an easy way by considering two partial cases:
when one contracts horizontal lines, i.e. in the limit
$a_1,a_3,a_4,a_6,a_8\to 0$, or vertical lines, i.e. at
$a_2,a_5,a_7,a_9,a_{10}\to 0$.
In both cases, one obtains recursively one-loop integrals which
can be evaluated in terms of gamma functions for general $\ep$.
On the other hand, taking such limits reduces to calculating
residues in some integration variables.
Consider, for example, the limit $a_5\to 0$.
We have $\Gm(a_5)$ in the
denominator but also the product $\Gm(a_5+z_1+z_4)\Gm(-z_1)\Gm(-z_4)$
which is singular in this limit. To reveal the singularity we
take residues at $z_1=0$ and $z_4=0$ and
obtain a factor $\Gm(a_5)$ so that the limit becomes nontrivial.

In the second step, one resolves the singularity structure in $\ep$,
taking residues and shifting contours, with the goal to obtain a
sum of integrals where one can expand integrands in Laurent series in
$\ep$. One can apply two strategies formulated in
\cite{MB1} and \cite{MB2}.
According to the first strategy, one performs an analysis of the
integrand to reveal how poles in $\ep$ arise. The guiding
principle is that the product $\Gm(a+z)\Gm(b-z)$,
where $a$ and $b$ can depend on the rest of the integration
variables, generates, due to
the integration over $z$, the singularity of the type
$\Gm(a+b)$. So, one thinks of integrations in various orders and
then identifies some `key' gamma functions which are crucial
for the generation of poles in $\ep$. Then one
takes residues and shifts contours, starting from first poles of
these key gamma functions. For contributions of the residues,
the same analysis and procedure
is applied. (See \cite{S4} for details.)

Within the second strategy \cite{MB2}, one chooses an initial value
of $\ep$ and values of
the real parts of the integration variables, $z_i,w,\ldots$ in
such a way that one can integrate over straight lines.
Then one tends $\ep$ to zero and whenever
the real part of the argument
of some gamma function vanishes one crosses this pole and
adds a corresponding residue which has one integration less and
is treated as the initial integral within the same procedure.

The third step of the method is to evaluate integrals expanded
in $\ep$ after the second step. Here one
can use  Barnes lemmas and their
corollaries to perform some of the MB integrations explicitly.
In the last integrations which usually carry dependence
on the masses and kinematic invariants,
one closes contour in the complex plane
and sums up corresponding series.

\section{Applying Gr\"obner bases to solve IBP relations}

Let ${\cal A} $ be
the  ring of polynomials of $n$ variables $x_1,\ldots,x_n$
and $\cal I\subset \cal A$ be an
ideal with a basis $\{f_1,f_2,\ldots,f_k\}$\footnote{
A ring is a set with  multiplication and addition.
A subset $\cal I$ of a ring $R$
is called a left (right) ideal if
({\em i}) for any $a,b\in \cal I$ one has $a+b\in
\cal I$ and ({\em ii}) for any $a\in {\cal I}, c\in R$ one has
$c a\in \cal I$ ($a c\in \cal I$ respectively).
A subset $\{f_i\}$ of $\cal I$  is  a \textit{basis} if
any $g\in \cal I$ equals
$\sum r_i f_i$ for some
$r_i\in {\cal A}$.}.
A basis is a {\em Gr\"obner basis}
if any polynomial $g\in \cal I$ is {\em reduced
modulo} this basis to zero for any sequence of reductions.
To define  {\em reduction}  one needs
an {\em ordering of monomials} $c x_1^{i_1}\ldots x_n^{i_n}$.
For the {\em
lexicographical} ordering,
$(i_1,\ldots,i_n)\succ (j_1,\ldots,j_n)$ (first monomial is
{\em higher} than the second one)
if there is $l\leq n$ such
that $i_1=j_1$, $i_2=j_2$, \ldots, $i_{l-1}=j_{l-1}$ and
$i_l>j_l$.
Then the {\em leading term} $\hat{g}$ of a polynomial $g$ is
the monomial which is higher than
any other monomial.
Now, the reduction is defined as follows.
Suppose that the
leading  term of a given  $g$ is divisible by the
leading  term or some $f_i$, i.e.
$\hat{g}=Q \hat{f_i}$.
Let $g_1=g- Q f_i$.
The leading  term of $g_1$ is lower than the
leading  term of $g$ and $g_1\in {\cal I}$ if and
only if $g\in {\cal I}$.
One can continue and proceed with $g_1$ as with $g$
and obtain similarly $g_2,g_3,\ldots$. The
procedure is repeated until one obtains $g_l\equiv 0$
or an element $g_l$ such that $\hat{g_l}$ is not divisible by any
leading  term $\hat{f_i}$. One says that $g$ is
reduced to $g_l$ modulo the basis $\{f_1,f_2,\ldots,f_k\}$.

If a given basis is not a GB one can construct a GB
starting from it and
using the {\em Buchberger algorithm}.
Suppose that $\hat{f_i}=w q_i$ and  $\hat{f_j}=w q_j$ where
$w, q_i$ and $q_j$ are monomials and $w$ is not a constant.
Define $S(f_i,f_j)= f_i q_j- f_j q_i\,.$
Reduce it modulo $\{f_i\}$ as described
above. If the reduction gives a non-zero polynomial add it to the
initial basis as $f_{k+1}$.
Consider then such $S$-polynomials for other pairs of elements
(including the new element)
and reduce them modulo the `current' basis.
If there is nothing to do one obtains a GB.
It has been proven by
Buchberger \cite{Buchberger} that such procedure stops after a
finite number of steps.

A classical problem is to find out whether a given element $g\in
{\cal A}$ is a member of $\cal I$ or not.
This problem  can be solved by choosing an ordering and constructing the
corresponding GB
with the help of the Buchberger algorithm.
After that, one applies the reduction procedure modulo the constructed
GB to verify whether a given element belongs to the
given ideal $\cal I$.
If $g$ is a monomial and $\{g_1,g_2,\ldots,g_l\}$ is a GB, then
\be
x_1^{a_1}\ldots x_n^{a_n}
=\sum_{i=1}^l r_i g_i +\sum c_{i_1,\ldots,i_n} x_1^{i_1}\ldots
x_n^{i_n}\;,
\label{GBdecomp}
\ee
where none of the monomials in the second sum is divisible by
highest terms of $g_i$. Since
$\{g_1,g_2,\ldots,g_l\}$ is a GB,
the last sum vanishes if and only if $g\in {\cal I}$.
If $g$ does not belong to $\cal I$, the last sum is non-zero, and
the part of $g$ belonging to $\cal I$ is completely included in
the first sum.

Let us now denote by $\cal I$ the
(left) ideal generated  by the elements $f_i$
which define IBP relations. To solve IBP relations is
to express the value of $F$ at an arbitrary point
$(a_1,a_2,\ldots,a_n)$ in terms of the values of $F$ in a few
specially chosen points, i.e. master integrals.
This problem can be solved similarly
to the algebraic problem described above.
Let us think of the case, where all the indices $a_i$ are
positive. Then
$F(a_1,a_2,\ldots,a_n)
=(Y_1^{a_1-1}\ldots Y_n^{a_n-1}\cdot F)(1,1,\ldots,1)\,.$
In this case it is reasonable to consider the operators
$Y_i$ as the main operators and get rid of the operators
$Y_i^{-1}$ by multiplying (of course, from the left) the operators
$f_i$ by sufficiently large powers of the operators $Y_i$.
Let us assume that we are dealing with such $f_i$ .

Let us observe that the situation is
quite similar to the above algebraic problem:
instead of polynomials in the variables
$x_1,\ldots,x_n$, we have polynomials in the shift operators
$Y_1,\ldots,Y_n$.
The natural idea is to turn, from the initial
basis $\{f_1,\ldots,f_l\}$, to a GB 
$\{g_1,\ldots,g_{l'}\}$.
Indeed, it is known that this can be done similarly to the above case:
one introduces an ordering and the notion of the highest term
which define the reduction modulo basis, then one can apply a
generalization of the Buchberger algorithm.
The motivation is the same: this is the GB
that characterizes the given
ideal in the `best' way, so that the parts which belong to the
given ideal do not belong to the second sum in (\ref{GBdecomp}).
Eventually, one obtains a similar relation,
$Y_1^{a_1-1}\ldots Y_n^{a_n-1}
=\sum_{i=1}^{l'} r_i g_i +\sum c_{i_1,\ldots,i_n} Y_1^{i_1-1}\ldots
Y_n^{i_n-1}\;,$
Let us apply it to $F$, take the value at $a_i=1$
and use the fact that the operators of $\cal I$
give zero on $F$. We obtain
$F(a_1,\ldots,a_n)
= \sum c_{i_1,\ldots,i_n} F(i_1,\ldots,i_n)$.
Integrals on the right-hand sides of
such relations (for various $a_1,\ldots,a_n$) are
master integrals.

However, any implementation of this similarity,
using a generalization of the
classical Buchberger algorithm, meets a lot of difficulties.
The given problem of solving IBP
relations becomes much more complicated at least
because one has to consider also non-positive indices $a_i$.

Another complication is the presence of the variables $a_i$
as non-commutative operators. Moreover, coefficients in monomials
in the shift operators, $Y_i$, can vanish at some
points.
Still let us imagine a situation where one can apply the
Buchberger algorithm to construct a generalization of
the GB
for solving a reduction problem in the case
of positive indices. Simplest examples show that the number of
the master
integrals associated with this region can be greater than the
number of the `true' master integrals.

These complications lead to the natural idea \cite{2S} to
change the strategy based on GB.
For a given family of integrals,
$F(a_1,\ldots,a_n)$, the whole region for each $a_i$
is decomposed into the regions with $a_i>0$ and $a_i\leq 0$.
The whole region of the multi-indices
is decomposed into $2^n$  {\em sectors}
$\sg_{\nu}=\{ (a_1,\ldots,a_n): a_i>0\;\;
\mbox{if}\;\; i\in \nu\,,\;\;
a_i\leq 0\;\;
\mbox{if}\;\; i \not\in \nu\}$
labelled by subsets
$\nu \subseteq \{1,\ldots,n\}$.
In a given  $\sg_{\nu}$ it is
natural to consider $Y_i$ for $i\in \nu$ and
$Y_i^{-1}$ for other $i$ as basic operators.
Another important point in extending Buchberger algorithm,
is taking into account
boundary conditions, i.e. specify the (\textit{trivial}) sectors where the
integrals vanish.

One has to construct \cite{2S}
a basis of Gr\"obner type for each non-trivial sector $\sg_{\nu}$.
The basic operations are the same as above,
i.e. calculating $S$-polynomials and
reducing them modulo current basis, with a chosen ordering.
The goal is to construct a so-called {\em sector} basis
\cite{2S} ($s$-basis)
which provides the possibility
of a reduction to master integrals \textit{and} integrals of
\textit{lower} sectors, i.e. $\sg_{\nu'}$ for
$\nu'\subset\nu$.
This point of the strategy is based on multiple examples
of solving IBP relations by hand, where one tried to reduce
indices to zero.
It turns out that, within this strategy, one
would construct a true GB only in the case of
the sector $\sg_{\emptyset}$ but
this sector is always trivial.

After constructing $s$-bases for all non-trivial sectors one
obtains a recursive (with respect to the sectors)
procedure to evaluate $F(a_1,\ldots,a_n)$ at any point and
thereby reduce a given integral to master integrals.
(See \cite{AVS} for details of the algorithm.)
Examples have shown that the algorithm
works at the level of modern calculations. In \cite{2S},
a non-trivial example of integrals with seven indices was
considered and, in \cite{GSS}, reduction problems for two families
of HQET integrals with nine indices were solved
successfully.

\section{Some perspectives}

For families of complicated Feynman integrals, a reduction
procedure was considered obligatory.
On the other hand, it was clear that
at least the second strategy of resolving singularities
in $\ep$ within multiple MB representations could be
formulated algorithmically and implemented on a
computer.
(I believe that the first strategy can also be automated
and that the two strategies can be combined in order to achieve an optimization 
of calculations.)
Recently, two algorithmic formulations have appeared
\cite{AnDa,Czakon}, so that, in complicated situations where
one fails to solve the reduction problem by Laporta's algorithm,
or by Baikov's method, or by using GB,
one can try to calculate every integral, at least numerically,
using these formulations.

In fact, the second of these algorithms \cite{Czakon} has been already
implemented in {\it Mathematica}.
This code provides a very good precision.
For the tennis court integral  discussed above
this numerical integration provides excellent agreement with the
analytic result of \cite{BDS}. (If such algorithm existed a
year ago, the authors of \cite{BDS} would be satisfied by this
powerful check and would not calculate asymptotic behavior
when $s/t \to 0$ by expansion by regions \cite{BS} :-) )
Anyway, evaluating
complicated  Feynman integrals, without
reduction, becomes now a reliable alternative.

\end{document}